\documentclass[draft]{agujournal2019}
\usepackage{url} %this package should fix any errors with URLs in refs.
\usepackage{lineno}
\usepackage[finalnew]{trackchanges} %for better track changes. finalnew option will compile document with changes incorporated.
\usepackage{soul}
\usepackage{color}
\draftfalse

\journalname{Geophysical Research Letters}

\usepackage{amsmath}

\begin{document}

%% ------------------------------------------------------------------------ %%
%  Title
%
% (A title should be specific, informative, and brief. Use
% abbreviations only if they are defined in the abstract. Titles that
% start with general keywords then specific terms are optimized in
% searches)
%
%% ------------------------------------------------------------------------ %%

% Example: \title{This is a test title}

\title{Statistical properties of magnetic structures and energy dissipation during turbulent reconnection in the Earth's magnetotail}

%% ------------------------------------------------------------------------ %%
%
%  AUTHORS AND AFFILIATIONS
%
%% ------------------------------------------------------------------------ %%

% Authors are individuals who have significantly contributed to the
% research and preparation of the article. Group authors are allowed, if
% each author in the group is separately identified in an appendix.)

% List authors by first name or initial followed by last name and
% separated by commas. Use \affil{} to number affiliations, and
% \thanks{} for author notes.
% Additional author notes should be indicated with \thanks{} (for
% example, for current addresses).

% Example: \authors{A. B. Author\affil{1}\thanks{Current address, Antartica}, B. C. Author\affil{2,3}, and D. E.
% Author\affil{3,4}\thanks{Also funded by Monsanto.}}

\authors{K. Bergstedt\affil{1,2}, H. Ji\affil{1,2}, J. Jara-Almonte\affil{2}, J. Yoo\affil{2}, R. E. Ergun\affil{3,4}, L.-J. Chen\affil{5}}

\authors{\today}
% \affiliation{1}{First Affiliation}
% \affiliation{2}{Second Affiliation}
% \affiliation{3}{Third Affiliation}
% \affiliation{4}{Fourth Affiliation}

\affiliation{1}{Department of Astrophysical Sciences, Princeton University, Princeton, New Jersey, USA}
\affiliation{2}{Princeton Plasma Physics Laboratory, Princeton, New Jersey, USA}
\affiliation{3}{Department of Astrophysical and Planetary Sciences, University of Colorado Boulder, Boulder, Colorado, USA}
\affiliation{4}{Laboratory of
Atmospheric and Space Sciences, University of Colorado Boulder, Boulder, Colorado, USA}
\affiliation{5}{NASA, Goddard Space Flight Center, Greenbelt, Maryland, USA}
%(repeat as many times as is necessary)

%% Corresponding Author:
% Corresponding author mailing address and e-mail address:

% (include name and email addresses of the corresponding author.  More
% than one corresponding author is allowed in this LaTeX file and for
% publication; but only one corresponding author is allowed in our
% editorial system.)

% Example: \correspondingauthor{First and Last Name}{email@address.edu}

\correspondingauthor{K. Bergstedt }{kbergste@pppl.gov}

%% Keypoints, final entry on title page.

%  List up to three key points (at least one is required)
%  Key Points summarize the main points and conclusions of the article
%  Each must be 100 characters or less with no special characters or punctuation and must be complete sentences

% Example:
% \begin{keypoints}
% \item	List up to three key points (at least one is required)
% \item	Key Points summarize the main points and conclusions of the article
% \item	Each must be 100 characters or less with no special characters or punctuation and must be complete sentences
% \end{keypoints}

\begin{keypoints}
\item An automated method to locate and identify plasmoids and current sheets in turbulent magnetotail reconnection regions has been developed
\item Plasmoids in a region of turbulent magnetotail reconnection have a decaying exponential size distribution from sub-electron to ion scale
\item Plasmoids and current sheets are significant contributors to parallel particle energization, but not to overall particle energization
\end{keypoints}

%% ------------------------------------------------------------------------ %%
%
%  ABSTRACT and PLAIN LANGUAGE SUMMARY
%
% A good Abstract will begin with a short description of the problem
% being addressed, briefly describe the new data or analyses, then
% briefly states the main conclusion(s) and how they are supported and
% uncertainties.

% The Plain Language Summary should be written for a broad audience,
% including journalists and the science-interested public, that will not have 
% a background in your field.
%
% A Plain Language Summary is required in GRL, JGR: Planets, JGR: Biogeosciences,
% JGR: Oceans, G-Cubed, Reviews of Geophysics, and JAMES.
% see http://sharingscience.agu.org/creating-plain-language-summary/)
%
%% ------------------------------------------------------------------------ %%

%% \begin{abstract} starts the second page

\begin{abstract}
    % GRL abstract limit is 150 words
    We present the first statistical study of magnetic structures and associated energy dissipation observed during a single period of turbulent magnetic reconnection, by using the \textit{in-situ} measurements of the Magnetospheric Multiscale mission in the Earth's magnetotail on July 26, 2017. The structures are selected by identifying a bipolar signature in the magnetic field and categorized as plasmoids or current sheets via an automated algorithm which examines current density and plasma flow. The size of the plasmoids forms a decaying exponential distribution ranging from sub-electron up to ion scales. The presence of substantial number of current sheets is consistent with a physical picture of dynamic production and merging of plasmoids during turbulent reconnection. The magnetic structures are locations of significant energy dissipation via electric field parallel to the local magnetic field, while dissipation via perpendicular electric field dominates outside of the structures. Significant energy also returns from particles to fields.
\end{abstract}

\section*{Plain Language Summary}
    Magnetic reconnection is an important mechanism for generating energetic particles in space and solar environments. Turbulent magnetic reconnection causes the development of many small-scale magnetic structures, such as locally helical or loop-like magnetic fields (plasmoids), or areas where oppositely directed magnetic fields are sandwiched together (current sheets). The exact formation and distribution of the structures, as well as the role the structures play in particle energization and the evolution of magnetic reconnection, is still unknown. Using data from the Magnetospheric Multiscale (MMS) mission, we developed an algorithm that is able to detect and identify the magnetic structures present in a region of turbulent magnetic reconnection. The number of structures was found to decrease with size as a decaying exponential, which is consistent with previous theories. The structures contributed strongly to the energization of particles parallel to the local magnetic field, but were not significant sites of energization overall. Overall energization is dominated by energization perpendicular to the local field outside of these structures. There is also significant energy return from particles to the fields.

%% ------------------------------------------------------------------------ %%
%
%  TEXT
%
%% ------------------------------------------------------------------------ %%

%%% Suggested section heads:
% \section{Introduction}
%
% The main text should start with an introduction. Except for short
% manuscripts (such as comments and replies), the text should be divided
% into sections, each with its own heading.

% Headings should be sentence fragments and do not begin with a
% lowercase letter or number. Examples of good headings are:

% \section{Materials and Methods}
% Here is text on Materials and Methods.
%
% \subsection{A descriptive heading about methods}
% More about Methods.
%
% \section{Data} (Or section title might be a descriptive heading about data)
%
% \section{Results} (Or section title might be a descriptive heading about the
% results)
%
% \section{Conclusions}

\section{Introduction}
    Magnetic reconnection is a process by which the topology of the magnetic field within a plasma is altered, allowing for the rapid conversion of magnetic energy into kinetic energy \cite{PARKER57}. It is responsible for the penetration of solar wind plasma into the magnetosphere \cite{RUSSELL78}, and plays an important role in powering solar flares and coronal mass ejections \cite{SWEET69,LIN00}. When reconnecting current sheets are sufficiently stretched to have large aspect ratios, plasmoids are expected to form via the tearing mode instability \cite{loureiro07}, leading to the multi-scale evolution of fast reconnection \cite<e.g.>{SHIBATA01,bhattacharjee09} across space and astrophysics including Earth's magnetotail \cite{JI11}. In the latter case, plasmoids have been observed via the ISEE-3 and GEOTAIL satellites over an extended period of time \cite{BAKER84,HONES84,RICHARDSON87,MOLDWIN92,NAGAI94,IEDA98,SLAVIN03} and more recently by the CLUSTER mission on the ion scales \cite<e.g.>{CHEN08_NPHYS,CHEN12}. Plasmoids are also routinely seen in kinetic simulations \cite<e.g.>{daughton06,DRAKE2006_SI}.  Therefore, a thorough analysis of the structures present in a reconnecting current sheet can shed light on the dynamics of fast reconnection, which in turn affect the global dynamics of the magnetosphere.
    
    An important feature of magnetic reconnection is the dissipation of magnetic energy to plasma particle energy through $\boldsymbol{J}\cdot \boldsymbol{E}$ where $\boldsymbol{J}$ and $\boldsymbol{E}$ are current density and electric field, respectively. There is an ongoing debate about whether the component of $\boldsymbol{J}\cdot \boldsymbol{E}$ along or across the local magnetic field, expressed as $J_{\parallel} E_{\parallel}$ and $\boldsymbol{J}_{\perp}\cdot \boldsymbol{E}_{\perp}$, respectively, is the primary source of particle energization \cite<e.g.>{DRAKE14,yamada18,fox18,PUCCI18}. Furthermore, whether the dissipation within the localized reconnection structures is significant~\cite<e.g.>{egedal12} or can be ignored~\cite<e.g.>{drake19} in a large system is still unclear.
    From the same MMS data used in this Letter, \citeA{ERGUN18} found that the main positive contributor to the overall $\boldsymbol{J}\cdot \boldsymbol{E}$ was $\boldsymbol{J}_{\perp}\cdot \boldsymbol{E}_{\perp}$ at frequencies at or below the ion cyclotron frequency, but did not examine the spatial correlation between energy dissipation and magnetic structures. Therefore, a detailed statistical study of magnetic dissipation, including the decomposition into parallel and perpendicular components within and outside of the magnetic structures can provide insight on these ongoing debates.
    
    %, $\sim$0.15 Hz, or a duration of $\sim$6 seconds. Electron-to-ion scale plasmoids, when moving sufficiently quickly e.g. in a reconnection jet, would be detected by spacecraft as higher frequency field perturbations. Therefore, we do not expect small-scale magnetic plasmoids or other structures to be a significant contributor to $\boldsymbol{J}_{\perp}\cdot \boldsymbol{E}_{\perp}$.  
    
    Many analytic and numerical studies have characterized possible size distributions of secondary islands in various regimes  \cite{UZDENSKY2010,FERMO10,FERMO11,LOUREIRO12,HUANG12,TAKAMOTO13,GUO13,LINGAM18,PETROPOLOU18}. Many of these studies have used Magnetohydrodynamic (MHD) models that are not generally applicable to kinetic scale plasmoids. However, the model developed by \citeA{FERMO10} is statistical in nature, and therefore can potentially be applied in a multiscale fashion. It postulates that plasmoids start small, then grow in size both by expansion and by plasmoid merging, leading to a smooth energy spectrum via an inverse-cascade \cite{NAKAMURA16}. A characteristic of the model of \citeA{FERMO10} is that for sufficiently large size (represented as a characteristic length scale), the number of plasmoids present in a reconnecting current sheet decreases exponentially with increasing plasmoid size. Studies have determined plasmoid size scalings in experimental plasmas \cite{DORFMAN14,OLSON16}, in solar plasmas via \textit{ex-situ} methods \cite{GUO13}, and in space plasmas via \textit{in-situ} methods \cite{FERMO11,VOGT14,AKHAVAN-TAFTI18}. \textit{In-situ} studies provide more detailed information on each plasmoid, but no \textit{in-situ} study thus far has utilized structures present in only a single turbulently reconnecting region. Plasma conditions varied considerably between each observation and  introduced unquantified uncertainties to the observed scaling. An analysis of the distribution of structures within a single turbulently reconnecting current sheet is desirable and necessary for accurately quantifying the plasmoid size scaling.

    The most common type of plasmoid observed in the magnetosphere is the flux rope, which is a helical magnetic field structure with a strong core field and an enhancement of the total magnetic field.
    Flux ropes have been extensively studied in space, and models of cylindrical force-free \cite{ELPHICRUSSEL83} and non-force-free \cite{LUNDQUIST50,LEPPING90} flux ropes are widely used. Flux ropes have been observed with complex internal structures \cite{STAWARZ18}, including enclosed waves \cite{WANG16} \add{and ongoing magnetic reconnection \protect \cite{oieroset16}}. Various other plasmoids have been observed in the magnetotail current sheet that do not have the typical cylindrical structure, including flattened flux ropes \cite{SUN19} and plasmoids which have loop-like field lines rather than helical \cite{ZHANG13}. These non-ideal plasmoids are indicative of the dynamic nature of magnetic reconnection. In a turbulent region, plasmoids may experience external forces which could slow or prevent their evolution into ideal cylindrical states. Therefore, for a turbulently reconnecting current sheet, in order to get a comprehensive survey of the plasmoids present, it is necessary to search for plasmoids that do not necessarily fit the ideal cylindrical flux rope model.
    
    Another question for a statistical survey of plasmoids is whether to identify the plasmoids `by eye', or to attempt an automated detection method. Automated methods are more rigorously defined and repeatable, and thus are less susceptible to human sources of bias. For example, methods have been developed to automatically detect flux ropes in satellite data \cite{SMITH17,Huang18}. These methods are repeatable, rigorous, and calculate valuable parameters such as the spacecraft's distance of closest approach to the center of the flux rope and the flux rope's radius. However, both methods are based on cylindrical flux rope models, force-free \cite{LUNDQUIST50} and non-force-free \cite{ELPHICRUSSEL83} respectively. These methods will not be suitable in a dynamic turbulent reconnection region which is likely to have large numbers of plasmoids which do not fit cylindrical flux rope models and unlikely to have obvious quiescent magnetic field backgrounds to compare the magnetic field fluctuations against. Therefore, an automated method has been developed to detect non-ideal plasmoids, as well as current sheets resulting from two different physical processes. This method has been used to probe the structure, dynamics, and dissipation of a turbulently reconnecting current sheet observed in the magnetotail.
    
\section{Observations and Methodology}

    MMS observed a period of turbulent reconnection on July 26, 2017 at 07:16:53 UT, and all four satellites collected about 17 minutes of data at their burst data rates. The electron energization and dissipation during this period was previously studied, and it was found that the main contributor to the overall net positive dissipation was due to $\boldsymbol{J}_{\perp}\cdot \boldsymbol{E}_{\perp}$ at or below the ion cyclotron frequency ($\sim$0.15 Hz in that region) \cite{ERGUN18}. It was additionally noted that there was a flux-rope-like structure that contained an exceptionally large $J_{\parallel} E_{\parallel}$, and that $J_{\parallel} E_{\parallel}$ was associated with electrons with energies up to 100 keV. Whether this finding can be generalized to the structures reported here was investigated. For this work, magnetic field data from the Fluxgate Magnetometer (FGM), which has a burst data rate of 128 Hz, is used \cite{RUSSELL16}. Electric field measurements from the axial and spin-plane double probes at a data rate of 8192 Hz \cite{Ergun2016,Lindqvist2016,Torbert2016}, and electron and ion moments from the Fast Plasma Instrument (FPI) are also used. FPI data is available at burst data resolution of 30 ms and 150 ms for electrons and ions respectively \cite{POLLOCK16}. \add{The region was characterized by depleted electron and ion densities of $<0.3 \textrm{ cm}^{-3}$, which led to typical ion and electron inertial lengths of $d_e \sim 10-40 \textrm{ km}$ and $d_i \sim 400-800 \textrm{ km}$ respectively.} In parts of the turbulent reconnection region, the electron density drops below 0.01 $\textrm{cm}^{-3}$, which means that the electron moments data will have large uncertainties during those time intervals. We elected not to use the electron moments data in these time intervals.

    \begin{figure}
        \centering
        \includegraphics[width = \textwidth]{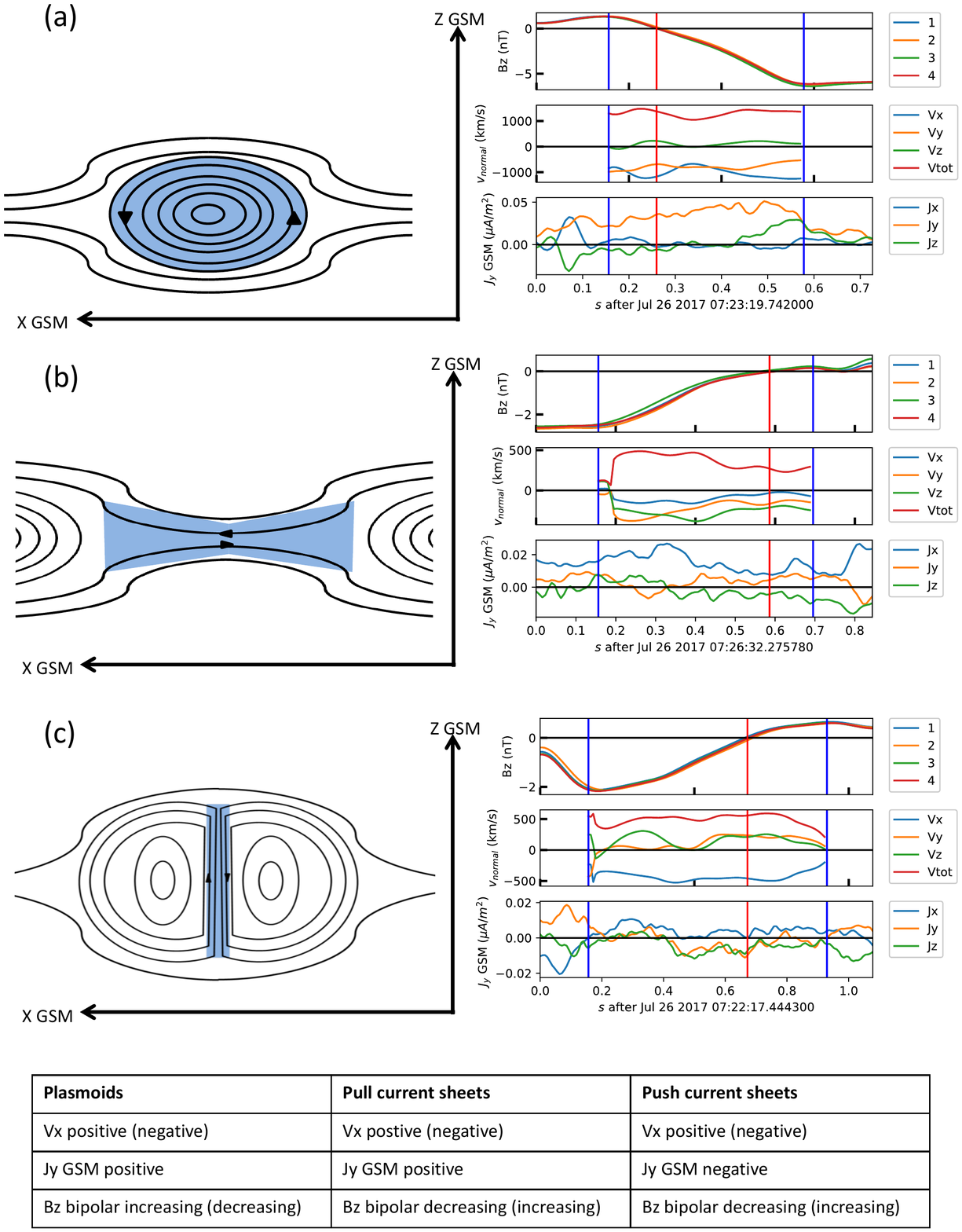}
        \caption{Left- cartoon model of the structure categories. Right- example of a structure from data. For (a) a tailward-moving plasmoid, (b) a tailward-moving pull current sheet,  (c) a tailward-moving push current sheet.
         The red vertical bar denotes the zero crossing and the blue bars denote the beginning and end of the structure. The magnetic field data was smoothed by a 6-point Hamming window for better estimation of the structure durations. \\
         Table: A summary of the selection criteria for the three structure categories. }
        \label{fig:Models}
    \end{figure}

    In order to categorize the structures as plasmoid\change{s}{-like} or current sheet\change{s}{-like}, idealized \add{two-dimensional} models of the structures and their orientations within the magnetotail current sheet were used, as shown in Figure \ref{fig:Models}. The magnetotail current sheet is generally in the X-Y GSM plane, with the current primarily in the +Y  ($J_y>0$) direction when sufficiently near the Y=0 plane. Plasmoids will generally be oriented with their invariant directions (e.g. the core direction of a flux rope) in the Y direction, and thus the currents within them will on average be in the +Y direction. We similarly assume that \lq\lq pull current sheets" ---current sheets between plasmoids that are not currently merging--- will maintain the same general orientation of the quiescent plasma sheet, and thus be approximately in the X-Y plane, with the current on average in the +Y direction. In contrast, \lq\lq push current sheets", which are current sheets formed by two plasmoids pushing into each other and potentially merging via reconnection, will be generally oriented in the Y-Z plane. The current direction is opposite that of our model plasmoids, pull current sheets, and the quiescent magnetosphere, so we expect currents within push current sheets to have components in the -Y direction ($J_y<0$). This distinguishes push current sheets from plasmoids and pull current sheets.
    
    In order to distinguish plasmoids from pull current sheets, we consider the direction of the bipolar signature. This direction will depend on the velocity that the structure is moving with respect to the spacecraft. The characteristic electron and ion speeds in this region are on the order of 100 km/s, whereas the MMS spacecraft are near their apogee and are therefore moving $<$10km/s. Therefore, we approximate that the MMS spacecrafts are stationary, and the relevant speed is that of the structures themselves. If the structure has motion in the +X (earthward) direction, a plasmoid will be detected by MMS as a first negative, then positive bipolar $B_z$ signature. In contrast, a pull current sheet will look like a first positive, then negative bipolar signature. If instead the structure is moving in the -X (tailward) direction, a plasmoid will appear as a first positive, the negative bipolar signature, while pull current sheets will appear as first negative, then positive bipolar signatures. Push current sheets will appear as positive-then-negative bipolar signatures if travelling in the +X direction, and negative-then-positive signatures if travelling in the -X direction. This leaves one category of structure without a known physical interpretation (structures with $J_y<0$ that have a negative-then-positive bipolar signature when travelling in the +X direction), but if the number of events tagged as this category are very small compared to those which have a known physical interpretation, it indicates that the approximate physical interpretations given to the other three categories are good approximations of the physical realities. Using this method, a bipolar $B_z$ structure in the data can be categorized as \change{a plasmoid, a pull current sheet or a push current sheet}{plasmoid-like, pull-current-sheet-like or push-current-sheet-like} via three considerations: 1) the direction of the bipolar signature (negative-to-positive or positive-to-negative), 2) the direction of the X component of the structure's velocity, and 3) the direction of the Y component of the current density. \add{Structures are hereafter referred to as 'plasmoids', 'pull current sheets' or 'push current sheets' depending on their categorization in this manner.} Examples of these three types of structures are shown on the right side of Figure \ref{fig:Models}, and a summary of the selection criteria is in the table at the bottom of that Figure.
    
    Structure candidates were first selected by identifying their bipolar $B_z$ signature in MMS1. In order to avoid some of the high-frequency transient turbulent magnetic field fluctuations, the data was first smoothed with a  six-point Hamming window. Upon finding potential structure candidates, their sizes were determined by finding the nearest local minimum in the negative part of the bipolar signature, and the nearest local maxima in the positive part of the bipolar signature. The number of comparison points for determining a local extrema was variable, but for the primary data 10 points to each side were used. At this point, if the other MMS satellites did not also observe a bipolar $B_z$ signature within the structure candidate, the structure candidate was discarded. The magnetic field data for the structure were then synced to a common timeline via a four-point Bartlett window \cite{ANALYSISMSC2}, and the lower resolution ion and electron moments data was synced to the same timeline via a cubic spline interpolation. The electric field data was synced to a common timeline via a linear interpolation to avoid artificial oscillations. 
    
    In order to calculate the structures' velocities and current densities, multi-spacecraft techniques were used. The current density was calculated via the curlometer technique \cite{ANALYSISMSC16,DUNLOP02}. The structure velocity was calculated in a two-step process. First, the dimensionality, invariant directions, and natural coordinates of the structure were calculated using the Minimum Directional Derivative (MDD) technique \cite{SHI05}, using the linear approximation of the magnetic field spatial gradient tensor from the barycentric coordinate approach \cite{ANALYSISMSC14}. Then the Spatio-Temporal difference (STD) method was applied to determine the velocity of the structure in its non-invariant directions \cite{SHI06}. The STD method cannot be used to determine the structure's velocity in its invariant directions (e.g. the core direction of an ideal flux rope), but motion in these directions by definition does not cause a large change of the magnetic field strength or direction. Therefore the velocity in the non-invariant directions is sufficient to determine the structure’s general motion in the X direction for categorization purposes.
    
    There are some additional limitations to the multi-spacecraft techniques used. For one, they are only reliable when all four spacecraft of the tetrahedron are within the same structure. The spacecraft spacing was $\sim$11km during this interval, while the electron skin depth was $\sim$15-20 km, so the multi-spacecraft techniques would be unreliable for sub-electron-scale structures. The tetrahedron also must have a Tetrahedron Quality Factor (TQF) of greater than 0.7 \cite{FUSELIER16}, which was satisfied during the interval. The techniques also assume approximate time stationarity, so any temporal fluctuations caused by the turbulence in the region could systematically affect the results from MDD and STD. The techniques also have some advantages; namely, they can be used at every data point in the time cadence, unlike other techniques that determine natural coordinates for structures such as minimum-variance analysis which can be used on data from a single spacecraft \cite{SONNERUP67,ANALYSISMSC8}. This also allowed us to evaluate the time-stationarity of the data by observing how the results from MDD and STD change with time throughout the structure.
    
\section{Statistical Results}
    \subsection{Magnetic Structure Properties}
    
        \begin{figure}
            \centering
            \includegraphics[width = \textwidth]{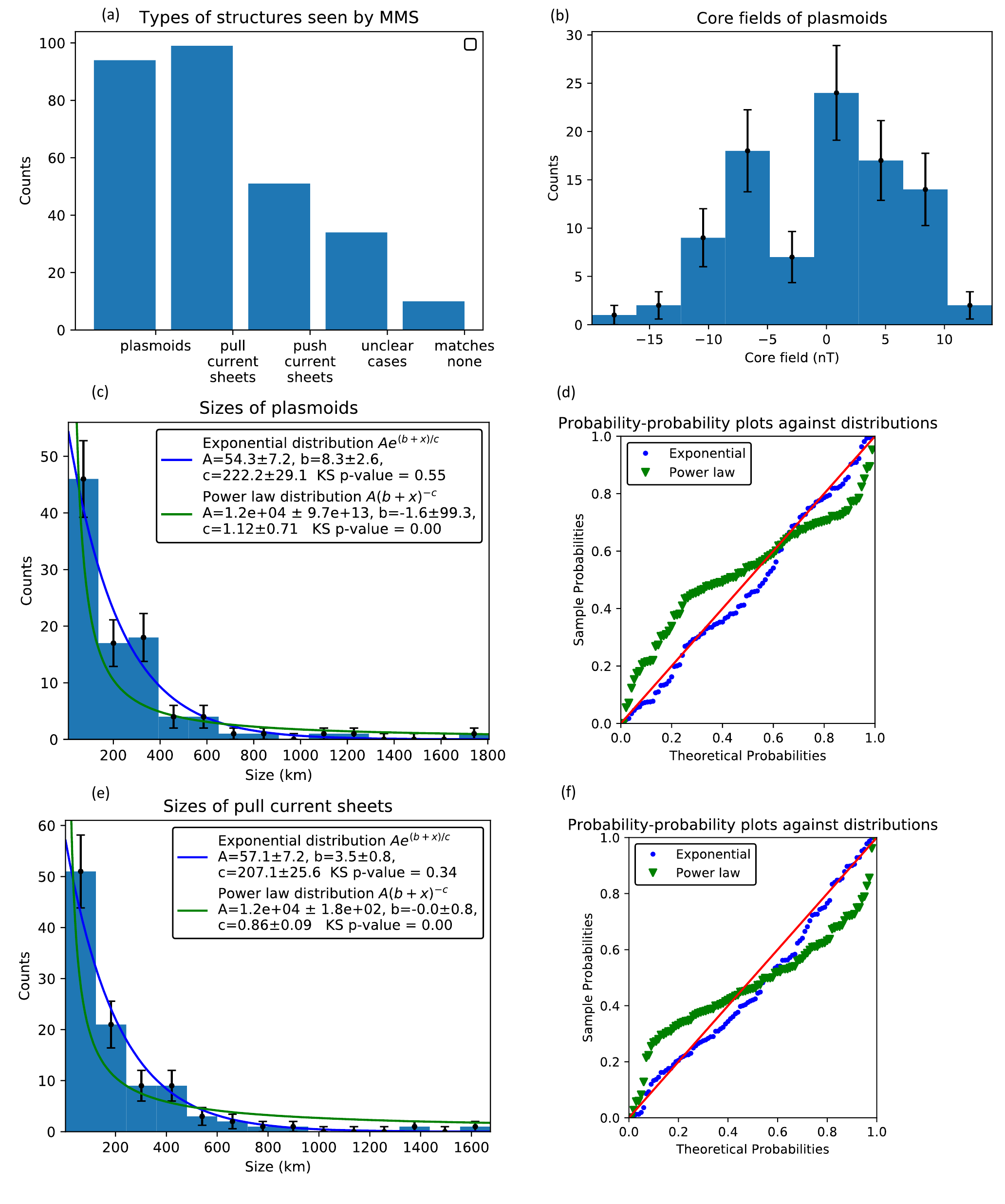}
            \caption{(a) Summary bar chart of structure types counted. (b) Histogram of the averaged core fields of the observed plasmoids. (c) Histogram of plasmoid sizes. (d) Probability-Probability plot of the exponential and power law fits for the plasmoid size data. (e) Histogram of pull current sheet sizes. (f) Probability-Probability plot of the exponential and power law fits for the pull current sheet size data.
            Error bars are from Poisson uncertainties. The errors on the fit parameters were computed by $n=100$ bootstrap using \cite{scikit-learn}. Kolmogorov-Smirnov tests \cite{KSBOOK} were performed on the exponential and power law fits, which accepted the exponential fit and rejected the power law fit.
            The probability-probability plots for the power law fits were done using a truncated power law distribution, as our selection mechanism is only sensitive to structures of a particular size range.
            
            \add{Typical electron and ion inertial lengths of $d_e \sim 10-40 \textrm{ km}$ and $d_i \sim 400-800 \textrm{ km}$ respectively show the given structures range from a few electron to a few ion inertial lengths in size.}}
            \label{fig:Hists_bar}
        \end{figure}
        
        There were 288 structures observed, and a summary of them is shown in Figure \ref{fig:Hists_bar}. Of these, 94 were plasmoids, 99 were pull current sheets, and 51 were push current sheets. 34 structures could not be categorized due to low certainty in the overall direction of the X component of the velocity or the Y component of the current density. 10 had sufficient certainty to be categorized, but did not match any of the given categories. These accounted for $\sim$3\% of the identified structures, so we conclude that the categories devised were adequate to describe the majority of sufficiently certain cases. Statistics were then performed on each of the structure types separately.
        
        An attempt was made to fit the observed plasmoids to force-free and non-force-free flux rope models in order to accurately measure the plasmoids' radii; however, the fits were inconclusive. Therefore, each structure's size was approximated by the product of the normal velocity of the structure and the duration of the structure. By this method, the majority of the structures were $<10d_{e}$ in size,  electron-scale. The size distribution histograms of plasmoids and pull current sheets are shown in Figure \ref{fig:Hists_bar}. The size data for plasmoids and pull current sheets were fit using maximum likelihood estimation (MLE), which does not require binning the data and is therefore more robust. Kolmogorov-Smirnov testing of the fits \cite{KSBOOK} found consistency with an exponential distribution but not with a power law. The push current sheets did not have a definitive fit (not shown). This decaying exponential is consistent with the prediction of Fermo et al. 2010 for sufficiently large scale size. This is the first \textit{in-situ} confirmation of Fermo et al.'s prediction from observations taken from a single turbulently reconnecting region. 
        
        Due to the turbulent nature of the magnetic field, we did not calculate the overall guide field of the reconnecting region. The guide field during magnetotail reconnection can change significantly on the timescale of less than a minute, so this event may not have a consistent overall guide field \cite{CHEN19}. Instead, we calculated the core fields of the observed plasmoids by finding the magnetic field strength along the most invariant direction determined from MDD analysis. This most invariant direction was generally primarily aligned with the GSM Y direction, so the distribution of the core fields of the plasmoids will be indicative of the total guide field of the region. Figure \ref{fig:Hists_bar} shows the distribution of observed core fields of the plasmoids, with positive core fields being aligned with the +Y GSM direction and negative core fields aligned with the -Y GSM direction. There are more plasmoids with positive core fields than negative ones, indicating a possible slight positive guide field. However, the core fields are not overwhelmingly in the +Y direction, indicating that the guide field was not strong, or was changing over the course of the event. In the case of stronger guide field, it would be possible to follow the procedure outlined by \citeA{NAKAMURA16} to use band-pass filtering to identify electron-scale flux ropes. However, for weak guide field the structures identified in this fashion may be the product of instabilities other than the tearing instability, and therefore the technique is not appropriate for this data.

    \subsection{Particle Energization and Dissipation}
        \begin{figure}
            \centering
            \includegraphics[width = \textwidth]{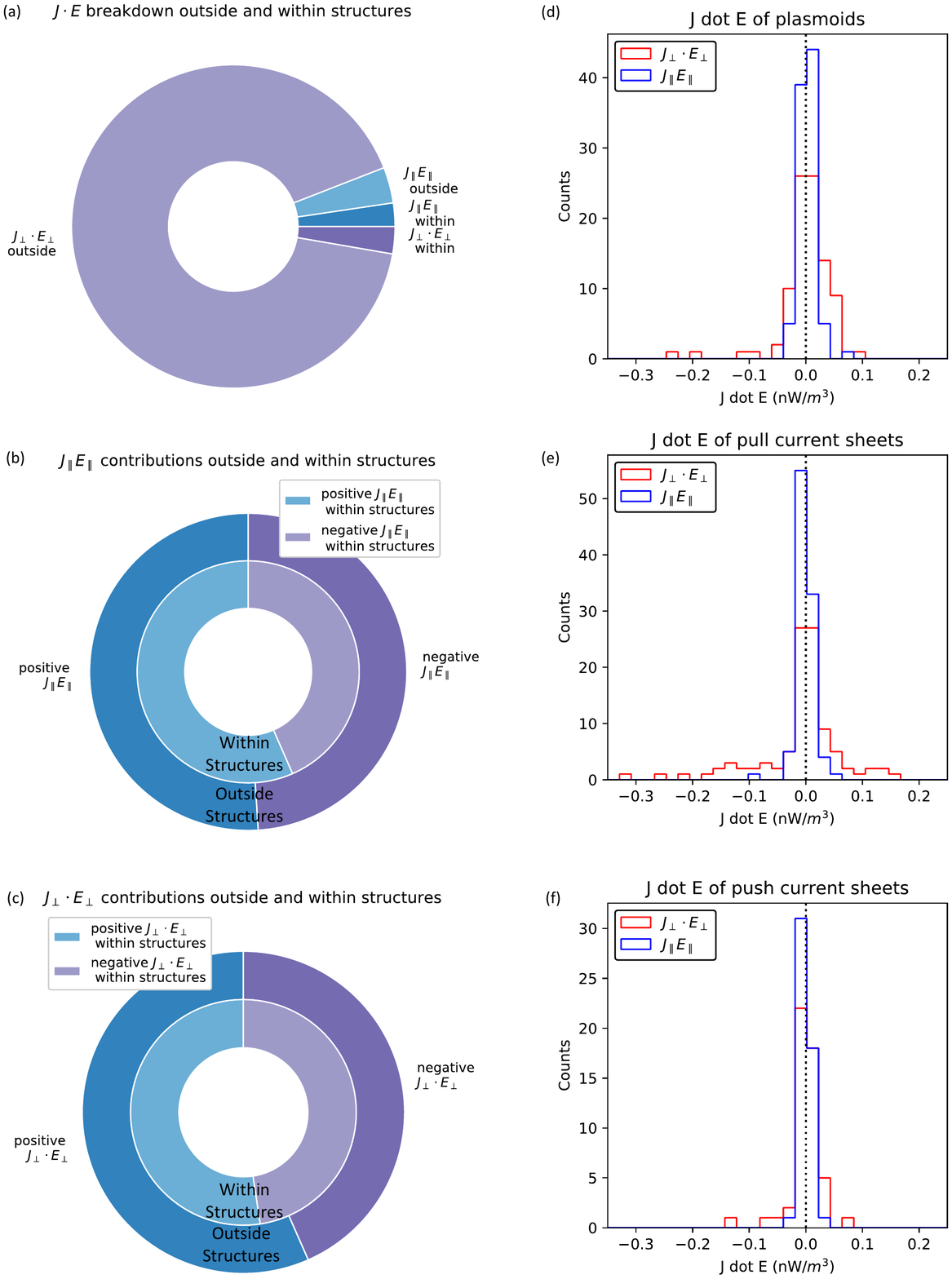}
            \caption{
            (a) Breakdown of the net contribution to $J_{\parallel} E_{\parallel}$ and $\boldsymbol{J}_{\perp}\cdot \boldsymbol{E}_{\perp}$ from the structures, compared to the regions outside the structures.
            (b) Comparison of the positive and negative contributions to $J_{\parallel} E_{\parallel}$. Outside circle is contributions from outside the structures, inside circle is contributions from the structures.
            (c) Comparison of the positive and negative contributions to $\boldsymbol{J}_{\perp}\cdot \boldsymbol{E}_{\perp}$. Outside circle is contributions from outside the structures, inside circle is contributions from the structures.
            (d-f) Histogram comparing the averaged contributions of magnetic structures to $J_{\parallel} E_{\parallel}$ and $\boldsymbol{J}_{\perp}\cdot \boldsymbol{E}_{\perp}$ for (d) plasmoids, (e) pull current sheets and (f) push current sheets.
            }
            \label{fig:J_Dot_E}
        \end{figure}
        
        To determine the dissipation mechanisms of the structures, we compared the $J_{\parallel} E_{\parallel}$ and $\boldsymbol{J}_{\perp}\cdot \boldsymbol{E}_{\perp}$ contributions from the structures and from outside the structures, summarized in the pie chart in Figure \ref{fig:J_Dot_E}. The structures covered $\sim$10\% of the total time duration of the region, but they contributed $\sim$40\% of the total $J_{\parallel} E_{\parallel}$ and only $\sim$3\% of the total $\boldsymbol{J}_{\perp}\cdot \boldsymbol{E}_{\perp}$. These electron-to-ion-scale structures are major contributors to the $J_{\parallel} E_{\parallel}$ in the region, which is consistent with \citeA{ERGUN18}'s identification of a flux-rope-like structure associated with large $J_{\parallel} E_{\parallel}$ and highly energized electrons of $>$100 keV. However, the breakdown between positive and negative contributions to $\boldsymbol{J}\cdot \boldsymbol{E}$ shows more complexity. As shown in Figure \ref{fig:J_Dot_E}, the regions both inside and outside of the structures have significant positive and negative contributions to $J_{\parallel} E_{\parallel}$ and $\boldsymbol{J}_{\perp}\cdot \boldsymbol{E}_{\perp}$. The structures have a larger ratio of positive to negative for $J_{\parallel} E_{\parallel}$, leading to their significant contribution to net $J_{\parallel} E_{\parallel}>0$. In contrast, the region outside of the structures has a larger ratio of postive to negative for $\boldsymbol{J}_{\perp}\cdot \boldsymbol{E}_{\perp}$, leading to a much smaller contribution from the structures, which are closer to parity. This breakdown shows that both within and outside of the structures there is ongoing energy conversion from fields to particles and vice versa, whereas the net energy exchange favors particle energization.
        
        The histograms of the averaged $J_{\parallel} E_{\parallel}$ and $\boldsymbol{J}_{\perp}\cdot \boldsymbol{E}_{\perp}$ are shown for the three major structure types in Figure \ref{fig:J_Dot_E}, and they confirm that the structures are sources of both positive and negative $\boldsymbol{J}\cdot \boldsymbol{E}$. The average perpendicular components have a larger spread than the parallel components by a factor of $\sim$2, indicating that $\boldsymbol{J}_{\perp}\cdot \boldsymbol{E}_{\perp}$ has the larger impact on overall $\boldsymbol{J}\cdot \boldsymbol{E}$, whether positive or negative. The histograms for the plasmoids show a bias towards positive $\boldsymbol{J}\cdot \boldsymbol{E}$, both for the parallel and perpendicular components, indicating these structures are on average sites of some particle acceleration. There are some notable outliers, but they do not significantly impact the structures' average contributions.
        
        Overall, $\boldsymbol{J}_{\perp}\cdot \boldsymbol{E}_{\perp}$ contributes $\sim$90\% of the total $\boldsymbol{J}\cdot \boldsymbol{E}$, whereas $J_{\parallel} E_{\parallel}$ only accounts for $\sim$10\%, and $\sim$85\% of the total average $\boldsymbol{J}\cdot \boldsymbol{E}$ comes from $\boldsymbol{J}_\perp \cdot \boldsymbol{E}_\perp$ outside of the structures. Therefore, the structures have a small contribution to the overall  $\boldsymbol{J}\cdot \boldsymbol{E}$, though some may serve as injection sites with large $J_{\parallel} E_{\parallel}$ which provide rapid energization to small populations of electrons, while the $\boldsymbol{J}_{\perp}\cdot \boldsymbol{E}_{\perp}$ between structures provides the largest net energization, such as proposed in \citeA{COMISSO19}. This result supports the use of codes which simulate particle energization during magnetic reconnection on larger-than-kinetic scales, such as the one detailed in \citeA{drake19}, but some handling of electron injection source terms may still be necessary.

\section{Conclusions and Discussion}
   
   We utilized \add{two-dimensional} models of the expected magnetic signatures of plasmoids, pull current sheets, and push current sheets to automate the detection and categorization of 288 magnetic structures within a 17-minute turbulent reconnection region. The majority of these had sizes between the electron and ion skin depths, making this the first statistical survey of mainly electron-scale structures within the same current sheet. It is possible to change the parameters of the detection algorithm to find systematically larger structures, but the focus of this work was on the smaller-scale ones, which may potentially be embedded within larger structures. 
   
   The estimated size distribution of the plasmoids was found to fit a decaying exponential, which is consistent with \citeA{FERMO10}'s statistical model of plasmoid distribution, growth, and merging. The presence of push current sheets consistent with plasmoid merging provides further evidence of the importance of merging plasmoid dynamics to the overall structure of the reconnecting current sheet. The bulk motion of the structures supports the analysis of \citeA{ERGUN18}, who observed a large-scale reconnection region with turbulent outflows. We also noticed that structure sizes were positively correlated with the structure speeds (not shown). However, the resolution limit of the magnetic field data prevents detection of small, fast-moving structures. Additionally, structure speed is used to calculate size, so there could be some artificial correlation. 
    
   The region was shown to have significant energy conversion from fields to particles and vice versa, but net $\boldsymbol{J}\cdot \boldsymbol{E}$ was positive for particle energization at the expense of the field energy. On average the structures were significant contributors to the net $J_{\parallel} E_{\parallel}$ of the region, contributing $\sim$40\% of the net $J_{\parallel} E_{\parallel}$. In contrast, 97\% of the $\boldsymbol{J}_{\perp}\cdot \boldsymbol{E}_{\perp}$ contribution was from the regions between the structures, meaning that these larger regions were the main contributor to the overall positive $J\cdot E$, which was comprised of 85\% $\boldsymbol{J}_{\perp}\cdot \boldsymbol{E}_{\perp}$ from outside of the structures. This is consistent with a model of the structures as injection sites, with strong localized $J_{\parallel} E_{\parallel}$ able to quickly accelerate electrons, which then can be slowly accelerated along with ions in the larger-scale regions of net positive $\boldsymbol{J}_{\perp}\cdot \boldsymbol{E}_{\perp}$. This indicates that the majority of the particle acceleration from these turbulent reconnection regions can be modeled using larger-scale physics, with the smaller-scale $J_{\parallel} E_{\parallel}$ injection sites largely ignored, or modeled as source terms of energetic electrons. Therefore, codes which are focused on capturing the larger-scale dynamics of reconnection regions (such as \change{Drake et al. 2019}{\protect \citeA{drake19}}), perhaps with added electron injection, should accurately describe the bulk of the particle energization in the reconnection region. 
   
    Fitting the plasmoids to mathematical models would yield more details about their structure. We found that the observed plasmoids did not fit the constraints of force-free or non-force free cylindrical models, but more general models were not tried. The use of other methods for ascertaining magnetic field topology, such as the first-order Taylor expansion method outlined in \citeA{FU15}, would also provide greater insight into the structure of this turbulently reconnecting region. 
    
    It would be valuable to repeat the analysis of this paper using a different plasmoid detection algorithm, such as the method detailed in \change{Nakamura et al. 2016}{\protect \citeA{NAKAMURA16}}, which requires strong guide field. A machine learning algorithm could possibly be more comprehensive than our algorithm, which has inflexible cutoffs for structure detection. This work did not explore whether the observed current sheets were reconnecting or not. If a nuanced automated method was developed to detect evidence of ongoing reconnection, additional information about the dynamics of the reconnection region could be obtained.
    
    \add{Another valuable expansion of this work would be to examine particular structures of interest from a three-dimensional viewpoint. Recent works such as \protect \citeA{oieroset16, OIEROSET19}} \add{have shown that structures which fit simple two-dimensional models such as that of a flux rope can have more complex three-dimensional topology, which can impact the onset and rate of reconnection. Given the large number of magnetic structures and potential for multiple X-line reconnection in this region, an in-depth three-dimensional exploration of even a few of the magnetic structures in this region has the potential to provide further insight into turbulent reconnection dynamics.}

\acknowledgments
This work was supported by the U.S. Department of
Energy’s Office of Fusion Energy Sciences under Contract No. DE-AC0209CH11466, by NASA under Grant No. NNH15AB29I, and by the National Science Foundation Graduate Research Fellowship under Grant No. DGE-2039656. Any opinions,  findings, and conclusions or recommendations expressed in this material are those of the authors and do not necessarily reflect the views of the funding organizations. Analysis scripts used for this manuscript can be found in the DataSpace of Princeton University (https://dataspace.princeton.edu/handle/88435/dsp01x920g025r). The data used is available from the MMS Science Data center (https://lasp.colorado.edu/mms/sdc/public/).

%% ------------------------------------------------------------------------ %%
%% References and Citations

%%%%%%%%%%%%%%%%%%%%%%%%%%%%%%%%%%%%%%%%%%%%%%%
%
% \bibliography{<name of your .bib file>} don't specify the file extension
%
% don't specify bibliographystyle
%%%%%%%%%%%%%%%%%%%%%%%%%%%%%%%%%%%%%%%%%%%%%%%

\bibliography{specifics,reconnection}

%Reference citation instructions and examples:
%
% Please use ONLY \cite and \citeA for reference citations.
% \cite for parenthetical references
% ...as shown in recent studies (Simpson et al., 2019)
% \citeA for in-text citations
% ...Simpson et al. (2019) have shown...
%
%
%...as shown by \citeA{jskilby}.
%...as shown by \citeA{lewin76}, \citeA{carson86}, \citeA{bartoldy02}, and \citeA{rinaldi03}.
%...has been shown \cite{jskilbye}.
%...has been shown \cite{lewin76,carson86,bartoldy02,rinaldi03}.
%... \cite <i.e.>[]{lewin76,carson86,bartoldy02,rinaldi03}.
%...has been shown by \cite <e.g.,>[and others]{lewin76}.
%
% apacite uses < > for prenotes and [ ] for postnotes
% DO NOT use other cite commands (e.g., \citet, \citep, \citeyear, \nocite, \citealp, etc.).
%

\end{document}